# Matching Trace Element Distribution to Mineralogical Phases in Ancient Biotechnology-Derived Metallic Salts: a Multimodal Analysis

**Sumera Rehman[1], Spyridon Panagiotopoulos[2], Georgios E. Christidis[3], Athena Athanassiadou[4], Alessandro Olivo[1], Effie Photos-Jones[5,*], Silvia Cipiccia[1,6*]**

[1]Department of Medical Physics and Biomedical Engineering, University College London, Gower Street, London WC1E 6BT, United Kingdom

[2]Department of Mining Engineering, University of Kentucky, Lexington, KY 40506, USA.

[3]School of Mineral Resources Engineering, Technical University of Crete, Chania, Greece

[4]Department of Prehistoric and Classical Antiquities and Museums, Ephorate of Antiquities of Pieria, Katerini, Greece

[5]School of Humanities, University of Glasgow, Glasgow, United Kingdom

[6]Diamond Light Source, Harwell Science and Innovation Campus, Fermi Ave, Didcot OX11 0DE, UK

* Corresponding authors: effie.photos-jones@glasgow.ac.uk, s.cipiccia@ucl.ac.uk.

## Abstract

Conventional X-ray fluorescence (XRF) and X-ray diffraction (XRD) analysis applied to the investigation of ancient metal salts used as pigments and/or therapeutics provide bulk average compositions in two stand-alone data sets; however, major elements aside, these two sets cannot inform on the spatial distribution of one with respect to the other.

To address this issue, we present here, first, a multimodal approach incorporating spatially resolved XRF, XRD and nanoscale X-ray imaging applied to the analysis of archaeological and experimental samples of synthetic lead carbonate ($PbCO_3$ - Greek *psimythion,* 5th-2nd century BCE); *psimythion* was used as a cosmetic and/or a therapeutic for external applications. It was produced according to a well-documented recipe dated to the 4th century BCE.

Second, we demonstrate that by using a multimodal approach we can confidently assign trace elements to individual crystalline phases and/or to highlight the existence of non-crystalline phases. Although the assignment of an element to a phase (i.e. its location) is now possible, how it arrived there (i.e. the mechanism) is not always clear. Our approach aims to highlight possible changes in the ancient artefact's chemical/mineralogical composition, across time, from its original manufacture, to use and burial, to excavation and conservation.

## Introduction

Natural and synthesized salts of lead, zinc, iron and copper have well documented long histories of use as both pigments and therapeutics. By the third millennium BCE there is evidence in Egypt of the manufacturing of synthesized lead chlorides, phosgenite ($Pb_2Cl_2CO_3$) and laurionite ($PbCl(OH)$); both black minerals were used as cosmetics around the eye as well as in lotions to prevent and treat eye infections[1-3]. In the Greek world of the 6th century BCE there is the first documentary evidence for a white lead salt, psimythion. A recipe for its manufacture is presented by Theophrastus two centuries later, in the 4th century BCE, involving the use of lead metal suspended over a fermenting liquid in a closed vessel [4] (see Supplementary Material, Note 1). Throughout the Greco-Roman world, psimythion was established as a well sought-after item of beautification, namely for the acquisition of a paler complexion both in body and face as well as an ingredient in wound treatment [5]. Variations of that recipe (Latin *cerussa*, *ceruse*) from the Roman period to well into the 17th century and as a cosmetic ensured that this material was produced in sufficient quantities to satisfy demand, with its deleterious/toxic effects on the skin of the wearer only gradually becoming fully appreciated. As part of medicinal (homeopathic medicine) preparations and for external applications only, it continued to be used until the late 19th century.

In a recent publication of ours we have proposed a mechanism underlying the 4th century BCE manufacture of lead carbonate, cerussite, which involves the action of microorganisms (specifically, their respiration products) on the lead plate [6]. In that paper the sequential transformation of the original lead plate first into lead hydroxide, then acetate then hydrocerussite and finally cerussite, has been proposed. We called this series of transformations, the preparation stage, since it was driven by $CO_2$ formation derived from the respiration of microorganisms within the liquid. The recipe then called for the scraping off the powder (the salts mentioned above) forming on the surface of the metal plate and its subsequent grinding and mixing in water. We called that stage the processing stage, following which only the insoluble lead compounds (cerussite/hydrocerussite) would be retrieved and used, with the soluble ones discarded. The resulting, near pure, cerussite could then be formed into pellets after mounting in suitable moulds [6].

Analyses, by different investigators since the early 1930s, of pellets of psimythion recovered from largely female graves have shown that the pellets indeed consist of near pure cerussite with only minor/trace amounts of hydrocerussite present. This corroborates our observation that for the period between 5th - 2nd century BCE, the Theophrastus recipe (Supplementary Material, Note 1) must have been the one used for the workshop-based preparation of psimythion via a biotechnology process the full details of which remain to be elucidated. Given the burial context of the pellets, the assumption has been that the deceased intended to 'use' psimythion as an item of beautification in the afterlife.

In addition to psimythion in pellet form, we undertook analysis of psimythion as powder recovered from copper/bronze vials within male burials, presumed to be those of a doctor and dating to the 4th century BCE [7]. Analysis was carried out with both, conventional XRF and XRD, as well as via their synchrotron equivalents.

Conventional XRF/XRD techniques rely on the illumination of the whole sample with an X-ray beam and recording a single fluorescence spectrum or a single XRD pattern. This type of analysis, here referred as one-dimensional (1D) analysis, can give only a bulk average composition in two separate data sets, 1D-XRD and 1D-XRF. In this scenario, it is possible to confidently allocate only the major elements to their corresponding mineral phases. In the case of minor and trace elements, this task is more difficult and open to speculation.

Extending the analysis from 1D to 2D, i.e. to a combined (XRF/XRD) spatially resolved analysis, is not possible in a laboratory with conventional benchtop XRF/XRD facilities or indeed in the field (with portable XRF-pXRF); this is due to the low flux of laboratory X-ray sources, which would make the process of analysis too time-consuming (several days). Spatially resolved elemental composition can indeed be obtained in the laboratory using scanning electron microscopy combined to energy dispersive X-ray spectroscopy (SEM-EDS) [8]. However, this type of analysis does not provide a simultaneous elemental composition with unambiguous identification of crystal structure.

Spatially resolved elemental/mineralogical analysis is easier to achieve in a synchrotron facility. With the availability of a much higher X-ray flux, it is possible to perform XRF and XRD analysis in 2D, on a specific plane, in a reasonable time scale (hours) and with micron resolution. The combined techniques have already been successfully applied to the analyses of pigments in art, and to the mapping of elemental and mineral composition at high spatial resolutions [7, 9-11]. Moreover, they can be combined to 2D X-ray ptychography [12] (XPty). This is a nanoscale X-ray imaging method, which measures the grain size and distribution of individual powder particles and can be particularly helpful in the insight it can provide into the sample's properties. One limitation of our technique is that-XRD mapping works well only when the grain size of the sample is smaller than the size of the X-ray illuminating beam [13].

In the aforementioned publication of ours [7] we addressed a specific question: how do the various trace elements featuring in the XRF analysis distribute within the different mineralogical phases obtained from XRD analysis? Given the use of these powders as cosmetics/therapeutics, it is important to be able to shed light on the origin of the trace elements within the samples and with some degree of confidence. There are several possibilities: a. trace elements in the psimythion powders could originate from the metal plate itself (see Theophrastus recipe in Supplementary Note 1); b. 'new' trace elements could be introduced during the processing stage to include metallic instruments (iron/bronze used to scrape off the powder) or the quartz-based grinding equipment (mortar/pestle); c. another 'set' of trace elements deriving from the burial environment (to include the container, if present, or the surrounding burial 'soil', if not) could have adhered or become embedded within the metal salt; and d. there could be any combination of the above.

In the above publication [7], we analysed two samples of psimythion: one was found within a copper container (sample 961) and the other within a bronze container (sample 964). The results showed that both samples consisted of mainly cerussite (>92%) with hydrocerussite as a secondary constituent. In the case of sample 964, quartz was also present.

In the case of sample 961 retrieved from its copper container, Fe and Cu were detected in trace amounts and did not distribute within the lead-rich phases; rather, they were found within a non-crystalline (organic?) phase which may have been part of the original pharmaceutical preparation in which psimythion formed a constituent part (organic content analysis pending).

In the case of sample 964, retrieved from the bronze container, Fe and Cu seem to be associated with quartz. Quartz grains could have been introduced in the mixture during grinding to fine powder and in a quartz-based mortar. On the other hand, Ag seems to distribute exclusively within the cerussite and hydrocerussite phases of the sample, suggesting that a silver-bearing Pb metal was used for the making of psimythion.

So, for these two samples the association of Fe and Cu with either an organic binder (961) or quartz particles (964) points to their 'arrival on the scene' as part of the processing stage, (for example, scraping the metal off the Pb plate and grinding it fine) while Ag was a trace/minor impurity in the Pb metal itself. It is interesting that for these two samples no Cu corrosion products (for example, malachite) deriving from the copper-based container were identified in the XRD analysis.

By building upon this previous work [7] and, using the same multimodal approach which integrates 2D-XRF, 2D-XRD, and X-ray ptychography, this paper progresses further in understanding the nature of manufacture of these metallic salts by examining three new archaeological samples (samples 4A, 9A and 13A, Figure 1) dated to the 4th century BCE. For a detailed background to their archaeological context see Supplementary Material, Note 2. Contrary to 961 and 964, the samples examined in the present study were not enclosed in copper-based vials but either in a ceramic (4A) or a lead-based container (9A). The third sample (13A) was found on its own, and not within a container, next to the body of the deceased.

In addition to the analyses of the archaeological samples, one experimental sample (32X) is presented here. It was prepared under field-based conditions and following the same 4th century BCE recipe as the archaeological samples (see Supplementary Material Note 1 and method section for further information on sample 32X). It should be pointed out here that the experimental sample was never ground, nor was it mixed with water to help remove possible water-soluble components (i.e. there was no processing stage). As such, the particle size was never guaranteed to be uniform, nor was the composition of the sample expected to consist of only the insoluble parts. As it turned out, 32X consisted of cerussite only (see discussion below).

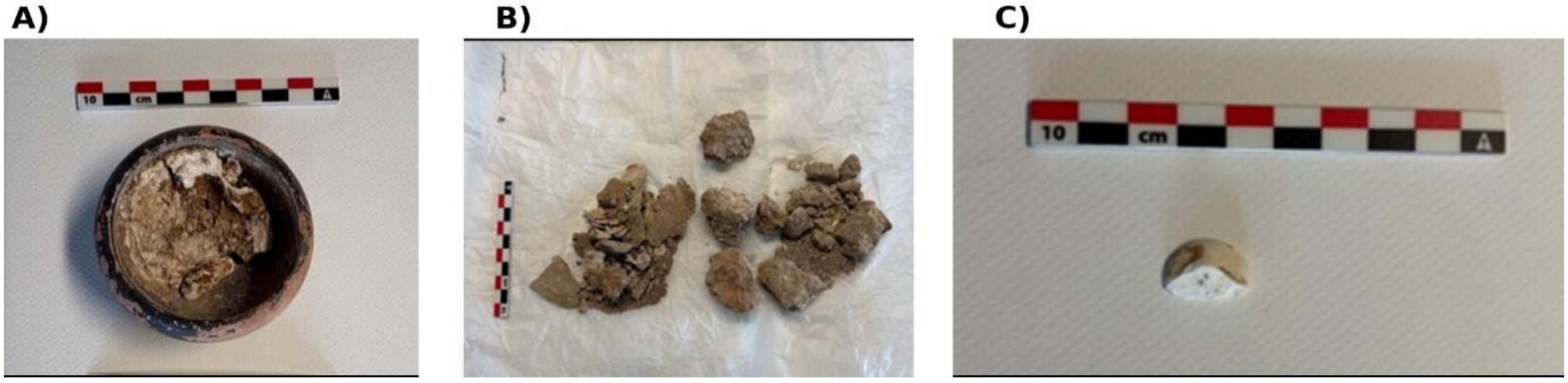

Figure 1: A) Ceramic bowl (skyfos) (PY7304) with psimythion (sample 4A); Alykes Kitrous, Pydna, S Cemetery (Tomb 68), Pieria, N Greece. Late 4$^{th}$ century BCE. B) Fragments of a lead vessel containing psimythion (sample 9A); Makriyalos, Pydna, N cemetery (Tomb 21, Plot 481. 2$^{nd}$ half of 4$^{th}$ century BCE. C) Fragment of pellet of psimythion outwith a container (sample 13A) Makriyalos, Pydna, N cemetery (Tomb 751, Plot 951). 2$^{nd}$ half of 4$^{th}$ century BCE. (Photos by I. Moschou)

**Method:**

*Samples:*

The three samples of psimythion dated to the 4th century BCE, retrieved from female burials include: a) a white powder, forming a residue within a ceramic vessel (Figure 1-A, sample 4A). This tomb belonged to a young woman buried with rich jewellery and items of beautification to include mirror, vases for ointments, and a ceramic vessel (PY7304) with psimythion in it; b) a white residue attached to the interior of a fragmented lead vessel (Figure 1-B, sample 9A). This tomb belonged to a young person with a small number of votive clay vessels and a lead vessel in fragments containing psimythion within; c) a fragment of a psimythion pellet found close to and outside a ceramic vessel (Figure 1-C, sample 13A). This tomb belonged to a female, and the small pellet was found at the foot of the deceased. It is considered that the tomb had been disturbed or looted prior to being excavated (see also Supplementary Material Note 2).

Experimental synthesis of psimythion was carried out following the same 4th century BCE recipe by Theophrastus (Supplementary Material Note 1), in which a lead metal plate was suspended over a liquid consisting of grape must (sample 32X) (Figure 2-A). Apart from occasional brief visual examinations, the lead plate and liquid were allowed to sit undisturbed in a closed container and over a period of approximatively two weeks. At the end of the two-week period the ‘fluff’ on the metal plate was scraped off with a knife and a subsample removed for analysis. Sample 32X is part of a series of still ongoing experiments using different starting materials to create a final product that closely resembles the archaeological psimythion [14]. The ancient recipe calls for *oxos,* a liquid commonly translated as vinegar [6] the exact nature of which, in terms of the microorganisms present, is, so far, relatively little researched. *Oxos*, as a drink, appears frequently in agricultural manuals of the ancient and Byzantine world [15] and indeed in the Bible [16].

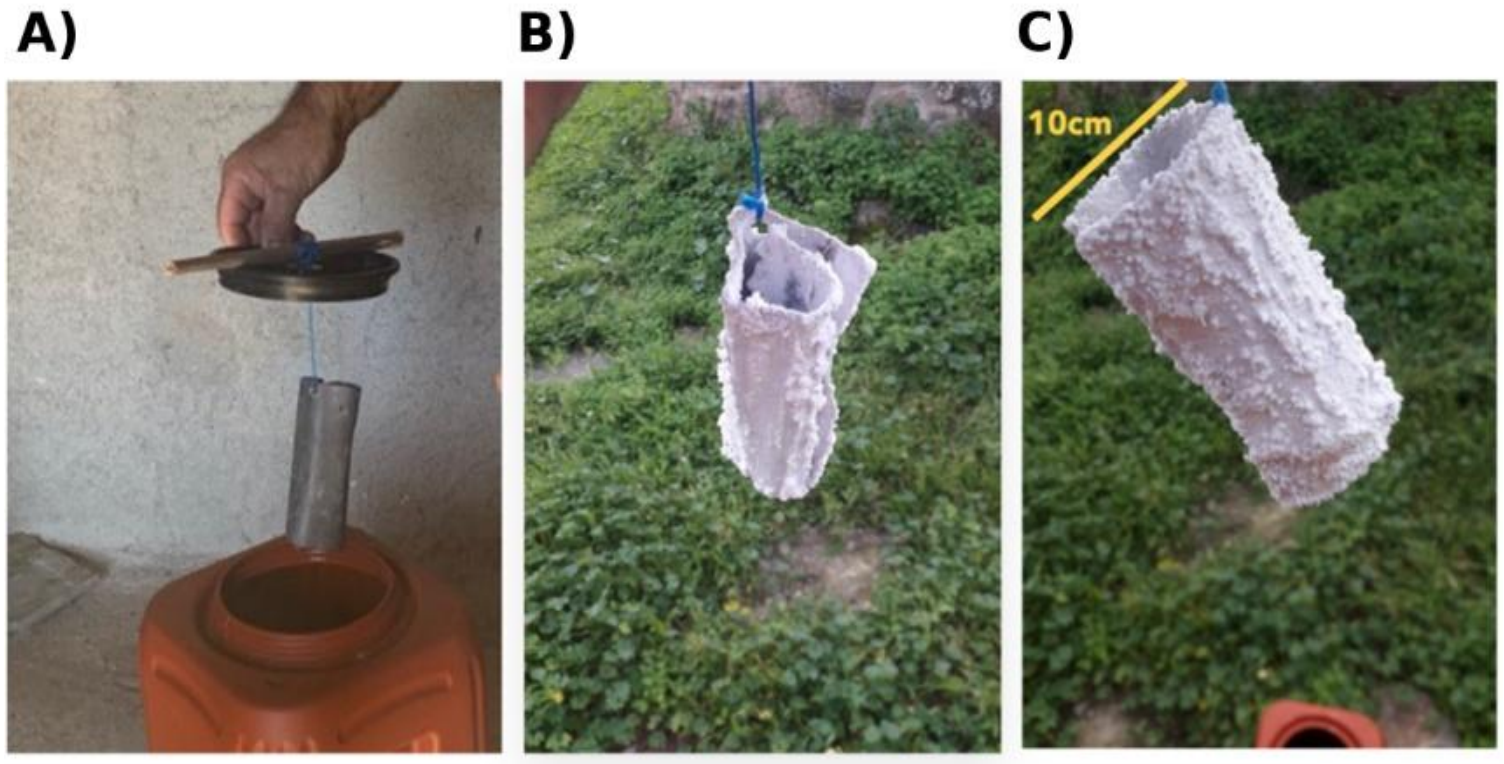

Figure 2: A) metal plate attached to the lid of a plastic container containing unfermented grape juice i.e. must. Figures 1B and 1C (middle and right) show the white powder which formed gradually on the surface of the lead plate (sample 32X) and over a period of 2-3 weeks in a closed environment. The powder was scraped off and without any further treatment it was subjected to 2D XRF/XRD analysis.

*Multimodal imaging*

The multimodal imaging was conducted at the I13-1 beamline of Diamond Light Source [17]. Three techniques were applied: 2D-XRF, 2D-XRD, and X-ray ptychography (XPty). For all the three methods, 16 keV X-ray beam was used. The X-ray energy was selected with a double-crystal Si-111 monochromator available at the beamline. The X-ray beam was focused down to spot size of approximately 10 μm diameter at the sample using a Fresnel Zone Plate optics (FZP, 400 μm diameter, 200 nm outermost zone width). The sample was mounted on a three-axis PI-Mars nano positioning stage for precise scanning.

For all the three techniques used for the analysis, the sample was scanned across the X-ray beam, both horizontally and vertically, following a regular raster grid pattern. XPty, XRD and XRF were performed sequentially by switching configuration as illustrated in Figure 3.

For the experiment the samples, all in powder form, were spread on Kapton tape dots. The dots (diameter 7 mm) were fixed on the top of a steel pin and were glued to a metal stub before being mounted on the PI-Mars scanning motors.

*2D-XRF and 2D-XRD:*

For the XRF and the XRD experiment the scanning step size was 10 μm. At each scanning step the XRF fluorescence spectrum was detected at an angle of approximatively 80° to the incident beam using a pre-calibrated four-channel Vortex silicon drift detector. The detector sensor was placed approximately 5 cm away from the sample.

The XRD acquisition was performed with the same scanning modality. The diffraction patterns were recorder at each scanning step with a large active area photon counting detector, the ExcaliburRX-3M [18] (500 μm thick Si substrate, 55 μm pixel size, 1093 × 2069 pixels) placed along the X-ray beam path, 17 cm downstream of the sample. The direct un-scattered X-ray beam was blocked just before the detector to avoid damaging the detector, with a lead central stop. The XRD detector was calibrated using cerium dioxide (SRM 674) and MICA (SRM 675) NIST standard reference materials. During the experiment, due to geometrical constraints, the XRD detection was limited to a maximum angle equivalent to 2θ=30° at 8keV.

For both the XRF and the XRD acquisitions, after each scanning step, the scanning motor was stopped, and the signal was acquired for 1 s and 2.5 s respectively.

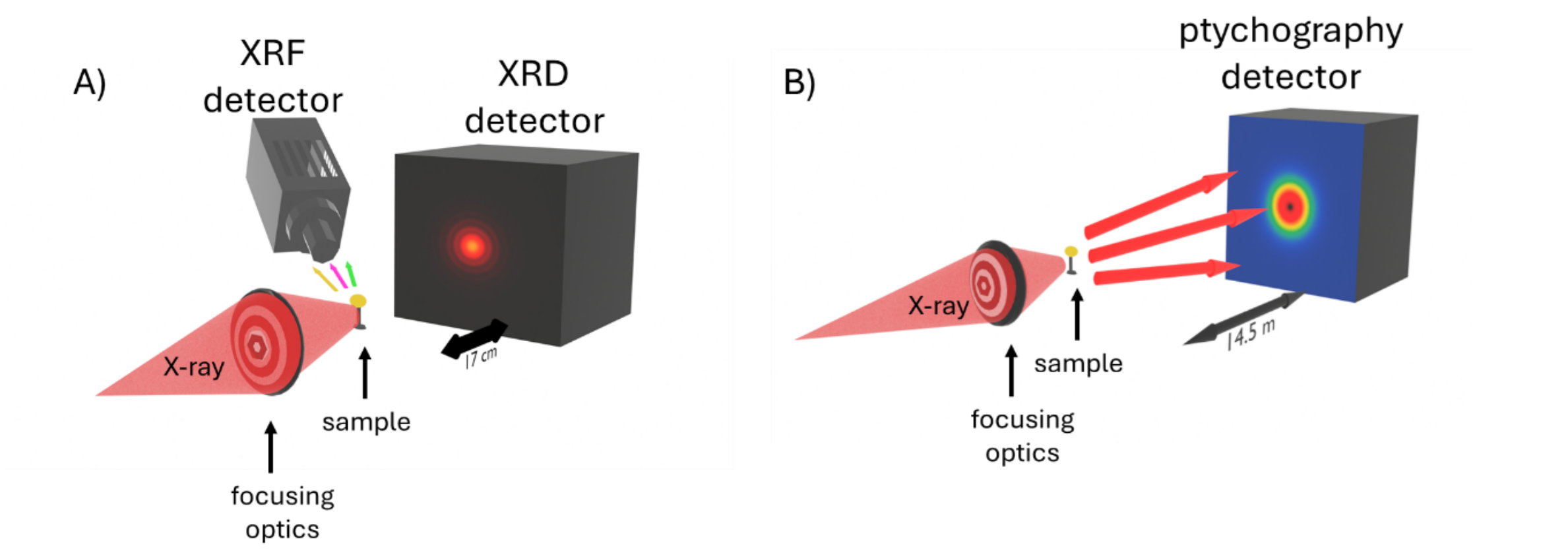

Figure 3: A) 2D XRD and 2D XRF setup. The sample is placed just after the focal plane of the FZP (not to scale). B) ptychography setup. Switching between setup A and B was achieved by moving the XRD detector out of the X-ray path to let the beam diffracted by the sample reach the detector in the far field.

*X-ray ptychography:*

XPty is a nanoscale scanning microscopy [19] widely employed at synchrotron facilities , with applications ranging from energy materials [20] to bioimaging [21]. A ptychographic acquisition consists of scanning the sample through a coherent probe beam of extent smaller than the sample, at overlapping steps while recording diffraction patterns in the far field. Unlike a standard microscope, there is no image forming lens after the sample. The lens is replaced with in an iterative algorithm [22-26] which inverts the diffraction patterns to retrieve the image of the sample: this renders the technique free from optics aberration. The method provides quantitative information of both the absorption properties of the sample, and the phase delay the X-rays experience in the sample. The phase delay is due to the change of X-ray refractive index between air and the sample, and it is directly proportional to the sample's electron density. This X-ray phase delay map gives access to fine structural information of the sample, down to few tens of nanometres in resolution.

During the XPty experiment, after the interaction with the sample, to reduce the noise coming from the air scattering, the diffracted X-ray beam propagates through a helium filled pipe to a photon counter detector (Eiger 500k photon-counting detector, 75 μm pixel size, 1030 × 514 pixels) placed 14.5 m downstream the sample (Figure 3-B). Unlike the XRF and XRD scans, the sample was moved continuously [27] across the illumination, to reduce the dead time due to the stop and start of the motors, which would have dominated the total acquisition time in this case, as for the XPty the exposure was reduced to 10ms per position. The motion followed regular raster grid trajectory and diffraction patterns were acquired at regular intervals in space of 0.5 μm and 1.0 μm in the horizontal and vertical directions respectively. To optimise the data collection and the image retrieval step by avoiding excessively large datasets, small scans of 100x100 $\mu m^2$ of neighbouring sample areas were performed, each scan was individually processed, and the reconstructions were stitched together to produce a larger picture of the sample.

*Data processing:*

The 2D-XRD and 2D-XRF data were processed using a Python script to perform background subtraction. For each of the samples under investigation, firstly the overall XRF emission spectrum was generated by summing the individual XRF spectra acquired at every scan position. Using the overall spectrum, the elements present in the sample were identified and their relative contribution to the sample composition was determined from the integrated areas of main peaks, after normalization to 100%. The same approach was used to generate an overall XRD pattern for the sample; this was used as input to the Bruker Diffrac.Suite.Eva software to identify the phases present in the sample. A semiquantitative estimation of their relative amount was obtained with Diffrac.Suite.Eva from the integrated areas of main peaks of the phases present, after normalization to 100 wt%. The precision is estimated to be ±10 wt% at the 10 wt%.

For each of the elements and phases identified, we produced a 2D distribution map where each point on the scanning grid corresponds to a pixel in the map. The value of each pixel is the number of counts recorded in the corresponding step of the scanning grid for the element and phase under consideration. Each pixel has a size of 10 μm, corresponding to the step size.

To determine whether a correlation exists between elements and phases and between trace elements and the main element (Pb), a pixel-by-pixel scatter plot was generated for each sample, a linear regression was computed, and the linear regression coefficient was calculated ($R^2$, see exemplar scatter plot of the cerussite phase versus the Pb signal in Figure 4-F). As the signal of the trace elements is weak, the elemental maps are noisy, and the statistical noise affect the results of the regression. To associate a level of confidence to the resulting $R^2$ the signal-to-noise ratio (SNR) of the elemental maps was used as a metric, defined as:

$$SNR = \frac{\mu}{\sigma}$$

Where $\mu$ is the average signal value calculated in a selected area of the XRF maps where a signal corresponding to the element of interest is detected, and $\sigma$ is the standard deviation of the noise, that is the standard deviation in an area of the map where no signal is detected. A high SNR value indicates high confidence and a low SNR value (in this work we set this to be <2) indicates low confidence of the correlation results.

The ptychographic image reconstruction was performed using 100 iterations of the ePIE algorithm [22], implemented within the PtyREX framework [28]. The pixel size of the ptychography images is 58 nm. The average particle size was determined from the ptychographic image using ImageJ [29]. The reconstructed ptychography maps were manually masked to exclude regions with missing tiles or poor reconstruction (e.g. at the edges of the scan where the beam overlap is reduced), thresholded to isolate particles, and analysed using ImageJ's automated particle analysis plugin. The resulting particle distributions were analysed in terms of their mean size and standard deviation.

**Results:**

For all the samples the following data were generated: a. XRF and XRD maps, b. the overall XRF spectrum and XRD pattern, and c. correlations of trace elements vs Pb and trace elements vs mineral phases. Considering the study is dealing with trace elements, and therefore with datasets affected by significant statistical noise, we expect the $R^2$ values to be generally significantly lower than 1. Therefore, in this study, based on visual inspections of the XRF and XRD maps, we set $R^2$=0.1 as the reference value between some correlation and low correlation between the investigated quantities.

*Experimental sample:*

For the purposes of this paper, we start by presenting the results from the experimental sample (32X), as a paradigm for trace element distribution deriving solely from metal composition and/or metal salt preparation (i.e. eliminating aspects of burial conditions originating either from the ground or the container).

The 2D-XRF maps for the sample 32X are shown in Figure 4-A for each of the elements detected on the overall XRF spectrum (Figure 4-B). The detected elements, their relative amount, and the SNR for each elemental map are summarised in Table 1.

Table 1: Sample 32X. Quantitative analysis of the XRF signal. The column ‘wt%’ shows the relative content of each element. The column “$R^2$(with Pb)” lists the linear regression coefficient for the distribution of each trace element versus the Pb signal. The SNR value is calculated for each map in Figure 4-A. The column “$R^2$(with cerussite)” lists the linear regression coefficient of the detected elements vs the single phase cerussite shown in sample 32X. The values without asterisks are calculated using the whole maps, those with the asterisks are calculated after excluding the top right corner where large grains are present

| | 32X | | | |
|---|---|---|---|---|
| | **wt%** | **$R^2$ (with Pb)** | **SNR** | **$R^2$ (with cerussite)** |
| **Pb** | 99.50 | -.- | 9.5 | 0.020/0.230* |
| **Cu** | 0.18 | 0.358 | 1.9 | 0.003/0.031* |
| **Ni** | 0.02 | 0.240 | 2.2 | 0.002/0.024* |
| **Ag** | 0.07 | 0.163 | 1.4 | 0.002/0.024* |
| **Fe** | 0.14 | 0.013 | 46.6 | 0.0002/0.002* |
| **Cr** | 0.06 | 0.181 | 1.9 | 0.002/0.022* |
| **Ca** | 0.03 | 0.056 | 22.7 | 0.004/0.045* |

The $R^2$ values are > 0.1 for most of the elements except for Fe and Ca where this is <0.1 (Table 1). The high SNR value for the Ca and Fe maps provides confidence on the $R^2$ results for these two elements, indicating that the distribution of Fe and Ca do not match the Pb distribution. This deduction is confirmed in Figure 4-A, where the ‘hot-spots’ in the Fe and Ca maps clearly do not match the Pb distribution.

The XRD analysis was performed on the same area as for the XRF (1x1 $mm^2$). The only identified mineralogical phase is cerussite (Figure 4-D, E). By calculating the linear regression of the cerussite signal versus the lead signal based on the whole map, unexpectedly, we obtain a very poor correlation ($R^2$~0.02). However, a careful inspection and comparison of the cerussite map with the Pb map, indicates that some of the large Pb particles do not produce any XRD signal (see black arrow pointing to one of these particles in Figure 4-D). This is most probably due to the large grains being single cerussite crystals in an unfavourable orientation to satisfy the Bragg condition and therefore, while they are recorded in the XRF map, they do not produce a signal in the XRD. If the area occupied by the large particles, at the top right and side of the cerussite map, is discarded, the correlation between the Pb and the XRD signal is much stronger than for the whole map, with an $R^2$ value of 0.23 (see Figure 4-F). By applying the same approach to the remaining elements, a similar increase of the $R^2$ is observed when a partial map is used instead of the whole one (Table 1, column "$R^2$ (with cerussite)").).

In addition, the $R^2$ is consistently <<0.1 for Fe and Ca, while their SNR value is >> 2. Therefore, we can assert with confidence that the two trace elements do not correlate with cerussite. For the remaining trace elements (Cu, Ni, Cr, Ag), the SNR is ≤ 2, suggesting that the confidence on the $R^2$ is low and as a result no definitive conclusions can be drawn regarding their association with cerussite.

The ptychography scan for the analysis of the powder structural properties was performed only on a 100 x 100 $\mu m^2$ area corresponding to the central area of the XRF map because of experimental issues. Due to the limited field of view imaged, no useful information can be extracted in terms of average particle size and its distribution for this sample.

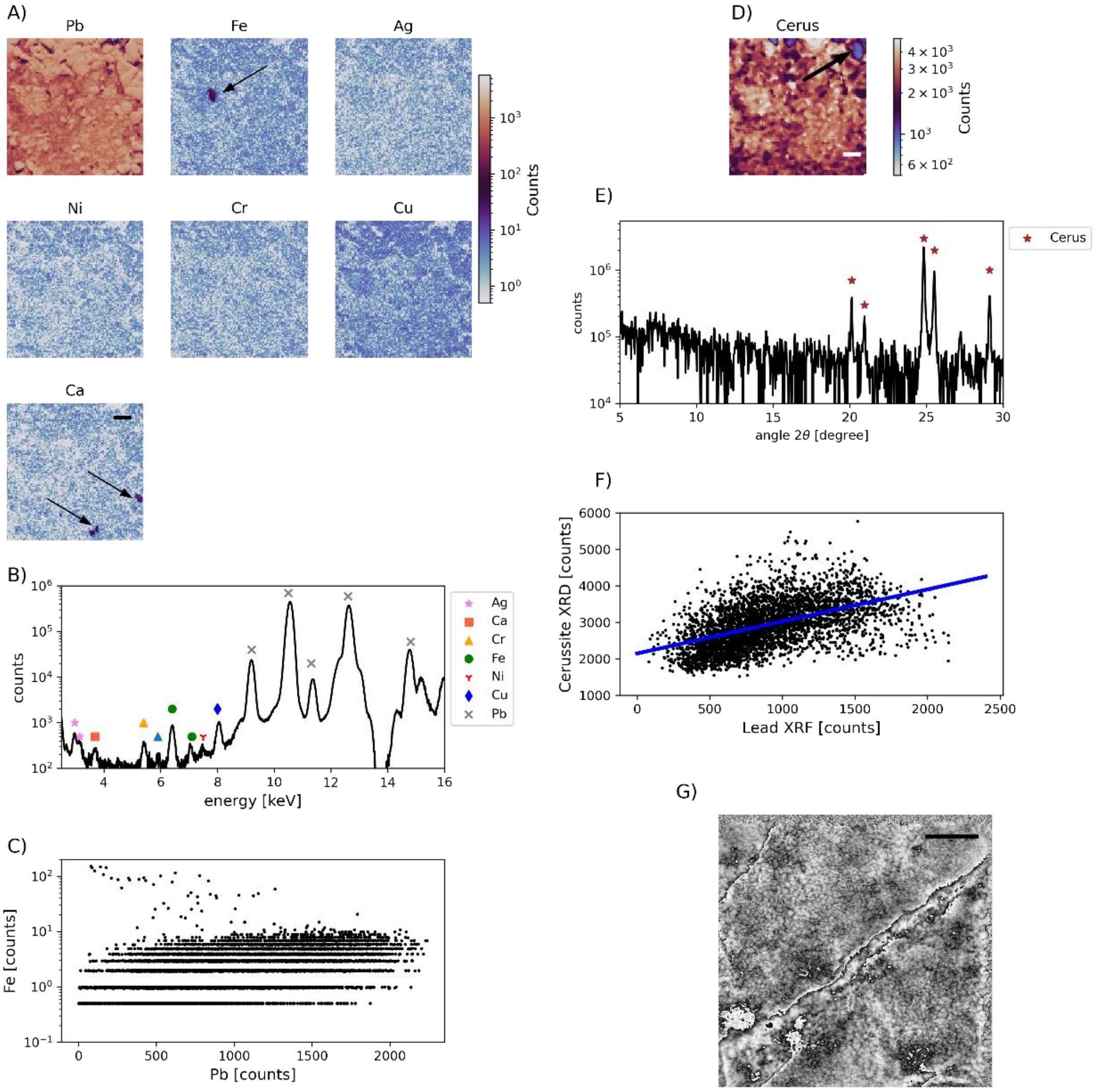

Figure 4: Sample 32X. A) elemental distribution map for Pb and all the trace elements (1x1 $mm^2$). The black arrows in the Fe and Ca maps indicate anomalous concentration of the elements with respect to Pb. The black scale bar in the Ca map indicates 100 microns and applies to all the images in A and D. B) total XRF spectrum obtained from the sum of each spectrum recorded during the scan. C) Fe distribution versus the Pb distribution. It should be noted that the outliers' high counts of Fe do not correspond to high counts of Pb. D) XRD map of cerussite, the only identified phase. The black arrow indicates one of the many anomalous large particles of Pb which do not show any signal in the XRD map. E) XRD pattern obtained as sum of all the pattern recorded during the scan. The $2\vartheta$ angle is calculated for Cu k-alpha energy. F) Cerussite counts versus the Pb counts neglecting the anomalous particles on the top and right side of the cerussite map, the XRD and XRF distribution show a linear trend. G) Ptychographic map of the central 100 x 100 $\mu m^2$ of the XRF map. The scale bar indicates 20 micron.

*Archaeological samples:*

The same approach used for the experimental sample was applied to the three archaeological samples, 4A, 9A and 13A. The measured elemental composition is reported in Table 2, together with the linear regression coefficient value of the trace elements vs the Pb, and the SNR of the elemental maps.

Table 2: Summary of the XRF analysis of the archaeological samples. Column ‘wt%’ lists relative composition. Columns ‘$R^2$ (with Pb)’ and ‘$R^2$ (with ph)’, shows regression coefficient of the linear regression of the detected element versus the Pb signal and the detected phase respectively. For sample 9A, as three phases are detected, we report the multiple regression element coefficient vs detected phases whose linear combination give the maximum $R^2$, using the following convention: the ‘*’ below the regression value indicate the best regression is obtained with a combination of calcite + cerussite; the ‘**’ below the regression value indicates the best regression is obtained with a combination of calcite, cerussite an quartz = calcite + cerussite + quartz.

| | | Pb | Cu | Cr | Fe | Ca | Ag | Ti | Ni | Zn |
|---|---|---|---|---|---|---|---|---|---|---|
| 4A | wt % | 97.83 | 0.32 | 0.26 | 0.97 | 0.14 | 0.35 | -.- | -.- | 0.13 |
| | $R^2$ (with Pb) | -.- | 0.280 | 0.095 | 0.087 | 0.203 | 0.116 | -.- | -.- | 0.290 |
| | SNR | 16.2 | 1.5 | 1.6 | 32.5 | 11.7 | 2.2 | -.- | -.- | 2.5 |
| | $R^2$ (with ph) | 0.382 | 0.118 | 0.035 | 0.065 | 0.227 | 0.028 | -.- | -.- | 0.128 |
| 9A | wt % | 93.98 | 0.26 | 0.14 | 4.46 | 0.78 | 0.17 | 0.21 | -.- | -.- |
| | $R^2$ (with Pb) | -.- | 0.648 | 0.351 | 0.010 | 0.001 | 0.347 | 0.007 | -.- | -.- |
| | SNR | 21.3 | 7.6 | 4.2 | 12.4 | 14.3 | 3.5 | 64.1 | -.- | -.- |
| | $R^2$ (with ph) | 0.640* | 0.377* | 0.193* | 0.021** | 0.015** | 0.191 * | 0.001** | -.- | -.- |
| 13A | wt % | 98.39 | 0.33 | 0.19 | 0.46 | 0.05 | 0.28 | -.- | 0.07 | 0.23 |
| | $R^2$ (with Pb) | -.- | 0.247 | 0.068 | 0.145 | 0.194 | 0.101 | -.- | 0.168 | 0.087 |
| | SNR | 6.5 | 1.8 | 1.1 | 7 | 2.4 | 0.8 | -.- | 1.8 | 2.3 |
| | $R^2$ (with ph) | 0.532 | 0.145 | 0.047 | 0.099 | 0.091 | 0.051 | -.- | 0.063 | 0.072 |

In sample 13A, regarding the trace elements detected with the XRF scan (see Figure 5-A, B), the Fe and Ca distributions match the Pb ($R^2$ > 0.1 and SNR >2), while Zn has a weaker correlation ($R^2$<0.1 and SNR>2). For the remaining trace elements, the calculated SNR is <2 and their relationship with Pb cannot be determined with confidence.

Only part of the 1x1 $mm^2$ area scanned for XRF was imaged in XRD mode (dotted area in Figure 5-A). The only identified phase is cerussite (see Table 5 and Figure 5-E), whose distribution map is shown in Figure 5-D, side by side with the corresponding area of the Pb map. As expected, the signal of the cerussite shows a good correlation with that of Pb ($R^2$>0.5). The Fe, Ca and Zn trace elements do not show a distribution strongly correlated with cerussite (SNR>2 and $R^2$<0.1). For the remaining elements, the high level of noise (low SNR) does not allow reaching a definitive conclusion. The ptychography scan was performed on the same area of the XRD (Figure 5-G).

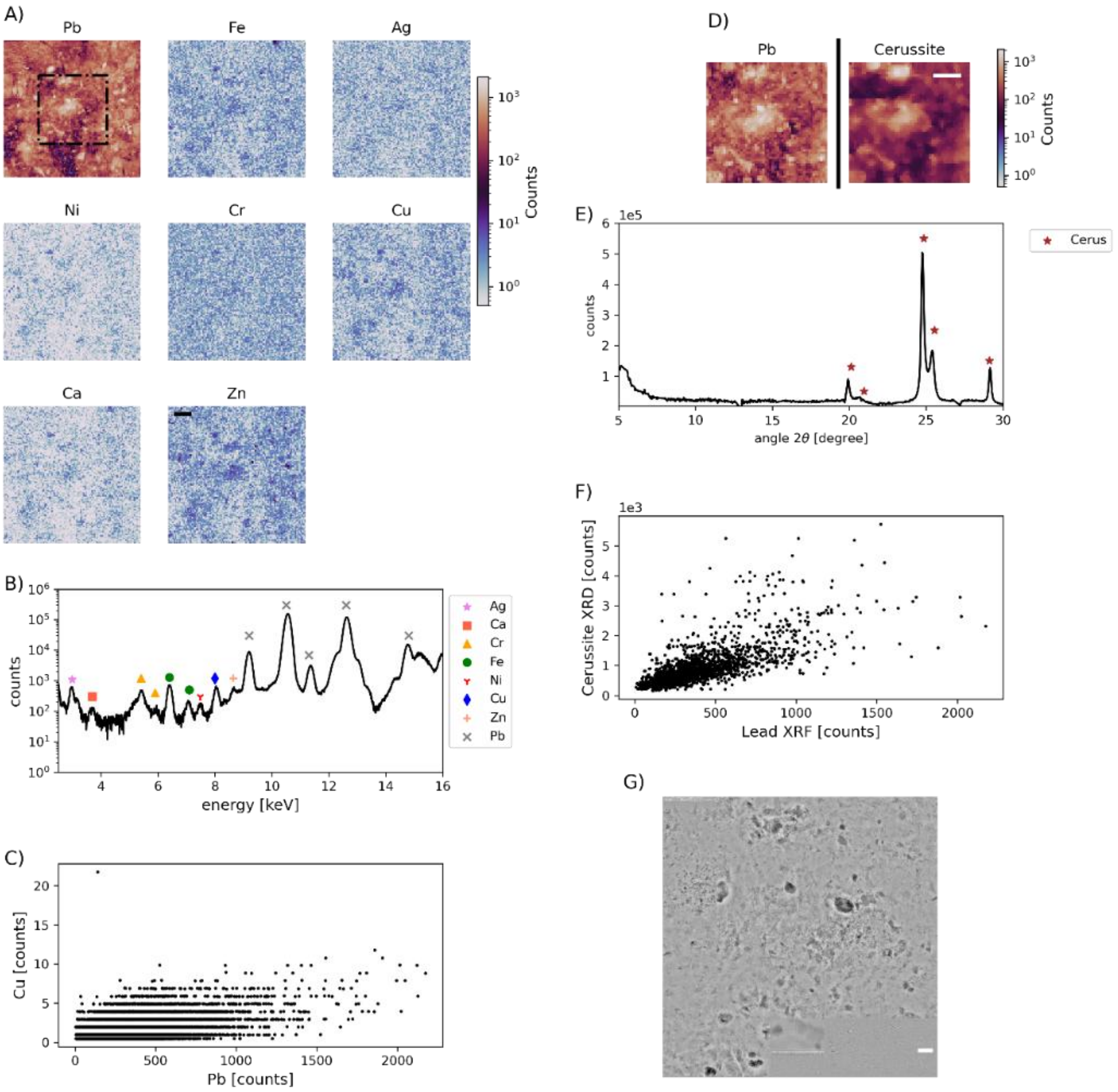

Figure 5: Sample 13A. A) elemental distribution map for Pb and all the trace elements (1x1 $mm^2$). The black dotted box indicates the area where the XRD scan was performed (0.5 x 0.5 $mm^2$). The black scale bar in the Zn map indicates 100 microns and applies to all the images in A. B) total XRF spectrum obtained from the sum of each spectrum recorded during the scan. C) Cu distribution versus the Pb distribution. D) XRD map of cerussite side by side with the corresponding Pb map area (the white scale bar corresponds to 100 microns). E) XRD patter obtained as sum of all the pattern recorded during the scan. The $2\vartheta$ angles correspond to the Cu k-alpha energy. F) Cerussite counts versus the Pb counts shows the linear trend. G) Ptychographic map of the central 500 x 500 $\mu m^2$ of the XRF map. The white scale bar indicates 20 micron.

The analysis of sample 4A is shown in Figure 6. The detected elements are listed in Table 2 and, according to the $R^2$ values validated with the SNR, the distribution of Ca, Ag and Zn can be related to Pb distribution with a good level of confidence (SNR >>2). The Fe map show hot spots that are not related to the Pb distribution, corresponding to a low $R^2$=0.065 and a high SNR. The XRD pattern shows the presence of only cerussite (Figure 6). Regarding the association of elements to phases, it follows, with good level of confidence, that Pb, as expected, is strongly correlated to cerussite, along with Ca and Zn, while Fe is not (Table 2).

Due to experimental time constraints, a smaller area (0.5 x 0.5 $mm^2$) was scanned for sample 9A, with respect to the other samples. The results of the analysis are presented in Figure 7. The list of the elements detected with the XRF is shown in Table 2. The $R^2$ values for the regression with Pb, strongly suggest that Cu and Ag match the Pb distribution, while Fe, Ca and Ti do not. Three main phases are identified in the sample, cerussite, quartz and calcite (Table 3). The cerussite and calcite maps are similar to the Pb distribution map (see regression coefficients in Table 2), unlike the quartz whose signal appears concentrated in a small area pointed to with an arrow in Figure 7-D. The ptychographic image of the same area is shown in Figure 7-G). Since in sample 9A three phases are detected, a multiple regression analysis was performed, instead of a simple linear regression of the elements versus a single phase; this is because each element could be associated with more than one phase. The phase combination which produces the highest $R^2$, together with the $R^2$ values, are listed in Table 2 (see Supplementary Material, Note 3, Table S1 for the complete set of regressions).

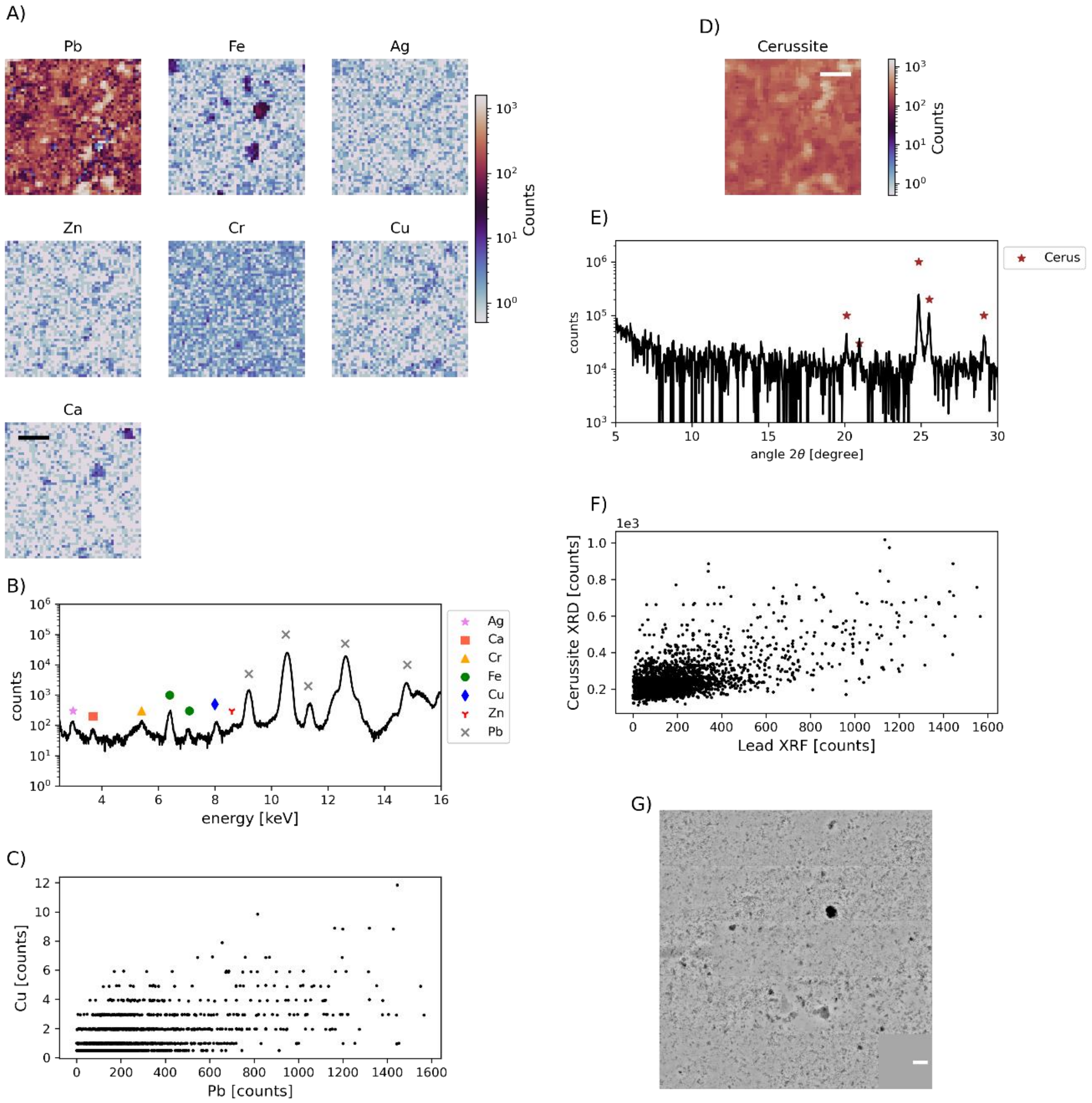

Figure 6: Sample 4A. A) elemental distribution map for Pb and all the trace elements (1x1 $mm^2$). The black scale bar in the Ca map indicates 100 microns and applies to all the images in A. B) total XRF spectrum obtained from the sum of each spectrum recorded during the scan. C) Cu distribution versus the Pb distribution. D) XRD map of cerussite (white scale bar 100 microns). E) The total XRD patterns showing that cerussite is the only identified phase. E) XRD patter obtained as sum of all the pattern recorded during the scan. The $2\vartheta$ angles correspond to the Cu k-alpha energy. F) The plot of the cerussite counts versus the Pb counts shows a linear trend. G) Ptychographic map of the central 500 x 500 $mm^2$ of the XRF map. The white scale bar indicates 20 micron.

Table 3: Archaeological samples: Relative percentage (wt%) of phases present.

| | 4A (wt%) | 9A (wt%) | 13A (wt%) |
|---|---|---|---|
| Cerussite | 100 | 69 | 100 |
| Quartz | -.- | 22 | -.- |
| Calcite | -.- | 9 | -.- |

Table 4: Average particle size and their standard deviation for the three archaeological samples.

| Sample | Mean ($\mu m^2$) | std ($\mu m^2$) |
|---|---|---|
| 4A | 0.30 | 2.41 |
| 9A | 0.72 | 5.1 |
| 13A | 0.15 | 0.97 |

Based on the multiple regression analysis it appears that Pb is associated both with calcite and cerussite. This does not mean that Pb is a component of calcite but that calcite and cerussite are well mixed together and their distribution overlaps. Similar conclusion with a good level of confidence can be drawn for Cu, Cr and Ag. Surprisingly the Ca distribution does not match the calcite distribution. This does not imply that there is no Ca in the calcite but that the most part of the XRF signal from the Ca either comes from an undetected phase (i.e. whose signal is outside the detection range of the system) or that it is not in crystalline form. Ti and Fe are not associated with any of the identified phases. The average particle size and size distribution of the archaeological samples based on the ptychographic imaging are summarised in Table 4.

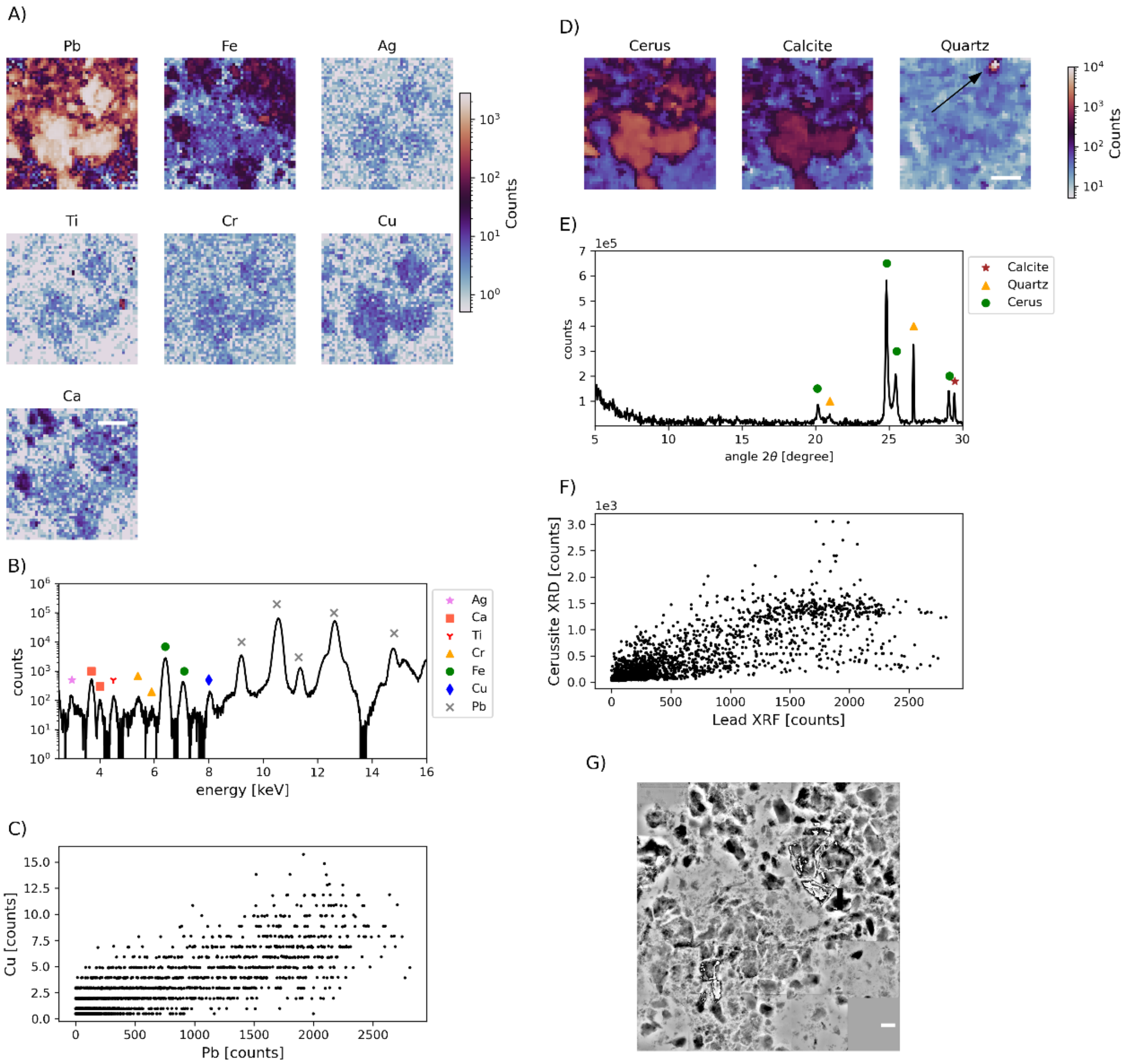

Figure 7: Sample 9A. A) elemental distribution map for Pb and all the trace elements (0.5 x 0.5 $mm^2$). The white scale bar in the Ca map indicates 100 microns and applies to all the images in A. B) total XRF spectrum obtained from the sum of each spectrum recorded during the scan. C) Cu distribution versus the Pb distribution. D) XRD map of the three phases identified in the sample (white scale bar in the quartz map corresponds to 100 microns). The black arrow points to the location of the quartz signal. E) XRD patter obtained as sum of all the pattern recorded during the scan. The $2\vartheta$ angles correspond to the Cu k-alpha energy. F) The plot of the cerussite counts versus the Pb counts shows a linear trend. G) Ptychographic map of the central 500 x 500 $mm^2$ of the XRF map. The white scale bar indicates 20 micron.

## Discussion

In this paper we present preliminary attempts at correlating trace elements with specific mineralogical phases within a metallic salt manufactured in antiquity and today via a biotechnological process [6]. Previously, this direct correlation between trace element and mineral was not possible, resulting in speculation; SEM-EDAX analysis can associate elements to phases, but the mineralogy of these phases is always approximated rather than confirmed. So, it has been common practice to attribute some trace elements to the original metal used, others to different stages of its preparation and yet others to 'contamination' while in the ground. This means that trace element allocation was carried out in an empirical and non-systematic way and on a 'best guess' basis. This correlation is important because the chemical and mineralogical profile of an ancient artifact cannot be thought of as fixed and unchangeable; and it is trace elements and their distribution across different phases that can shine light onto the artefact's 'life trajectory' from its manufacturing stage through use and burial, subsequent excavation and conservation.

As mentioned in the introduction we distinguish three stages in the production of the psimythion powder (or pellet) following a well-defined recipe. These include a) the procurement of the Pb plate; b) the production stage, i.e. the formation of metal salts through microbial respiration within the liquid(oxos); c) the processing stage i.e. scraping the ensuing powder off the metal plate, immersing it in water to remove soluble components and retaining only the insoluble precipitate i.e. cerussite ($PbCO_3$) and hydrocerussite ($Pb_3(CO_3)_2(OH)_2$); d). (optional) placing the powder in moulds in the shape of a pellet and allowing it to dry.

Experimental sample 32X was not subjected to stages c) and d), the processing and moulding stages respectively. From the archaeological samples, 13A was indeed pressed in a mould (Figure 1-C), while samples 961, 964, 4A and 9A were in powder form. Table 5 summarises key trace elements and their potential association with one or more of the three phases identified within the six samples discussed in this paper, namely, cerussite, calcite and quartz.

Table 5: Results of trace element to phase association for the six samples discussed here, including the two from [7]. We use the following symbols: '**X**' = not related to the phase; '**✓**' = related to the phase. We do not report the Ni as this was only detected in sample 9A and not associated to any phase.

| Sample | Phase | Fe | Cu | Ag | Zn | Cr | Ca | Ti |
|---|---|---|---|---|---|---|---|---|
| **32X** | cerussite | **X** | **X** | | | | **X** | |
| **4A** | cerussite | **X** | ✓ | | ✓ | | ✓ | |
| **9A** | cerussite | **X** | ✓ | ✓ | | ✓ | **X** | **X** |
| | calcite | **X** | ✓ | ✓ | | ✓ | **X** | **X** |
| | quartz | **X** | **X** | **X** | | **X** | **X** | **X** |
| **13A** | cerussite | **X** | ✓ | | **X** | | | |
| **961** | cerussite | **X** | **X** | | | | | |
| **964** | cerussite | **X** | **X** | ✓ | | | | |
| | quartz | ✓ | ✓ | **X** | | ✓ | ✓ | ✓ |

Starting with the experimental sample 32X, it is noted that Fe is never associated with cerussite; indeed, there is no crystalline phase that would 'host' Fe. Interestingly, and always with regards to Fe, the same observation holds true for all archaeological samples. Therefore, the possibility that Fe derives from the metal plate can be excluded. On the other hand, Cu seems, in some samples, to be associated with cerussite (4A, 9A, 13A) but not in others (32X, 961, 964). Similarly for Ca: in some samples, Ca is associated with cerussite (sample 4A) but not in others (samples 32X and 9A). The possibility that Fe or Cu derived from the tool (iron or bronze) used to scrape off the powder can certainly not be excluded. Ag, when detected, is always associated with cerussite (9A, 964) confirming that the original metal for these two samples had some, albeit trace, silver content. 'Cr when detected, is associated with cerussite (9A). Zn is found associated with cerussite in 4A, while the association is not clear in 13A (due to the $R^2$ being close to but lower than the 0.1 limit we adopted as threshold for the correlation). On the other hand, Ti shows no association with cerussite (9A). Mineralogically speaking Cr is less closely associated with Pb than with calcite. Cr's association with both cerussite and calcite (but not with quartz) in sample 9A suggests a possible substitution of Pb by Ca. The association of Ca with Cr is less clear.

Calcite, which as mentioned, is present only in sample 9A, is linked with Ag, Cu and Cr but not Fe, Zn or Ti (see Table 5). Quartz is present in 9A and 964; all elements, apart from Ag, are 'hosted' by quartz in 964 but the opposite is true for sample 9A (Table 5).

The above description which links elements to some mineralogical phases, but not others, appears to follow few overarching rules; instead, it appears to reflect the conditions pertaining to individual samples.

Cu shows a clear association with cerussite in samples 4A, 9A and 13A, but not in samples 32X, 961 and 964.The lack of association of Fe with cerussite, inclusive of the experimental sample 32X, raises the question of which phase actually 'hosts' Fe. That phase was invisible to XRD during our experiments. It is suggested that Fe may be in the form of an amorphous hydroxide/oxyhydroxide which might be trapped within the $PbCO_3$ matrix. Alternatively, it could be adsorbed on the surface of $PbCO_3$ [30, 31] or on organic matter. Searching for the non-crystalline phases within the sample is beyond the scope of the approach presented here, since the multimodal analysis rests on XRF/XRD, alone.

Some caveats need to be considered. First, the possible presence of an organic constituent (for example beeswax) in what must have constituted a therapeutic mixture for external applications (i.e. samples 961 and 964) needs to be investigated and, if applicable, to be considered. It is less likely that an organic binder would have been introduced in pellets like 13A [6]. Nevertheless, it is rather interesting how these pellets manage to retain their original shape; they are rarely seen in a crumbled state pointing to a surface cohesion, the nature of which remains to be investigated.

Second, all samples need to be ground to same particle size to eliminate large crystals creating difficulties in the analysis (see example in Figure 4, sample 32X). As mentioned earlier, the experimental sample 32X was never subjected to grinding and subsequent washing followed by filtering as prescribed in the original recipe.

Third, it is important to run multiple sub-samples from the same sample to establish a coherent and confident picture of each element's association with a particular phase. Sample inhomogeneity can be a defining parameter when reporting on the composition of archaeological materials, and it seems that, in the case of psimythion, inhomogeneity was considerably reduced by grinding and washing.

The association of Pb with both cerussite and calcite (sample 9A) and the lack of association of calcium with calcite (also in sample 9A), appears, at first surprising; nevertheless, this has been reported previously, in the case of calcite replacement by cerussite via the reaction by Pb aqueous solution [32, 33]. The process is controlled, in part, by sorption and precipitation reactions at the calcite surface [32] or by a dissolution-precipitation mechanism [33]. However, in the present study and with reference to experimental sample 32X calcite was not involved in the preparation of cerussite; the sample was not subjected to washing (where Ca-rich water may have bene involved), neither had the sample come at any point in contact with a calcium-rich source (a plastic container was used); finally, there was no subsequent burial of the sample. Yet Ca was present and not associated with cerussite. We are unable to explain its source.

The state of sample 9A, i.e. the cerussite contents of a lead vessel shown in an advanced state of decay (see Figure 1-B) needs also to be addressed; it may be that the 'soil' covering the powder may be composed of Ca-Pb phases forming a 'cake', rather than consisting of ordinary 'dirt'. Similarly in sample 4A, although no phases other than cerussite were identified there.

**Summary and Conclusions**

In this paper we argue that by using a multimodal approach we can confidently assign trace elements Ag, Zn and Cr to the Pb metal, while we can exclude the possibility of Fe being associated with the metal. On the other hand, we found that Cu may or may not be associated with Pb metal. We highlight the complex relationship between Pb and Ca so often seen in samples of psimythion (i.e. Ca in the pXRF, no calcite in the XRD) but we are presently unable to offer a definitive explanation for the mechanism of action.

Despite limitations in data interpretation, we suggest that our proposed approach goes a step further than what is currently available. Namely, it allows the investigator to delve deeper and with confidence into trace element origin and distribution within different phases. When discussing ancient artefacts, the term 'chemical/mineralogical composition' should perhaps be viewed as more 'fluid' than rigid. This is not simply on account of inhomogeneities inherent in the preparation of the artefact but also because of alterations introduced during the ancient artefact's trajectory across time. In succeeding to confidently assign elements to phases, some ambiguities, although perhaps not all, can be removed. Finally, we note that the multimodal approach can be applied onto other archaeological/historical materials, well beyond metallic Pb salts.

**Data availability**

Data generated and analysed from this experiment can be accessed on request by contacting the corresponding author.

**Acknowledgements**

We acknowledge Diamond Light Source for the provision of beamtime under proposal MG36736-1. We are grateful to Dr. Sarah Day (I11 beamline, Diamond Light Source) for kindly providing the MICA and $CeO_2$ standards used for X-ray diffraction calibration. We thank Dr. Darren Batey, Tim Arden, and Dr. Kudakwashe Jakata from I13-1 beamline for their invaluable support in the setup and run of the experiment. We also acknowledge the help of Mr Spyros Potsis, Aegina, Greece for the preparation of sample 32X and Mr Ioannis Moschou, Ephorate of Antiquities of Pieria, for the photographs in Figure 1.

**Author contributions**

E.P-J. and S.C. conceived the experiment. A.A. procured the samples. S.R. and S.P. were responsible for the sample preparation. Data collection was carried out by S.R., S.P., G. E.C., S.C. Data analysis and interpretation were performed by E.P-J, S.R., G.E.C., A.O., and S.C. The manuscript was written by S.R., S.C. G.E.C and E.P-J. All authors reviewed the article.

**Competing interests**

The authors declare no competing interests.

**Additional information**

The online version contains supplementary material available at TBC.

## Supplementary Material for: “Matching Trace Element Distribution to Mineralogical Phases in Ancient Biotechnology-Derived Metallic Salts: a Multimodal Analysis”

**Note 1:**

Theophrastus’ (*On Stones*, 56) recipe for *psimythion* making reads as follows:

‘lead (*molybdos*) about the size of a brick is placed in jars (*pithos*) over vinegar (*oxos*) and when this acquires a thick mass, which it generally does in ten days, then the jars are opened and a kind of mold (*evros)* is scraped off the lead, and this is done again until it is all used up. The part that is scraped off is ground in a mortar and decanted frequently, and what is finally left at the bottom is white lead *(psimythion*)’ (Caley and Richards 1956, 188)[4] (the transliterated Greek, in parentheses, is our addition.)

**Note 2:**

**Sample 4A**

Provenance: Pit Grave 68, Makriyialos, S Cemetery of Pydna, Chrysohoides Plot. Date of excavation: 2000. Excavators: M Bessios and Athena Athanassiadou.

A pit grave, oriented along a NW-SE axis. The walls of the grave were coated with plaster forming two distinct colour bands, red above and yellow below. It is a female burial.

The deceased was a young woman laid upon a wooden bier, with her skull oriented towards the southeast. She was accompanied by a number of items of jewellery: a. a diadem consisting of a lead band to which bronze spirals and gilded terracotta fruit-shaped ornaments were attached; b. a pair of gold earrings depicting Erotes; c. a gold necklace with a stone pendant; d. four gold beads on each shoulder; e. four rings, of gold and iron, on her left hand) and f. terracotta vessels (an oil lamp, a kantharos, two alabastra, a small amphora for unguents, and an unguent vessel); g. terracotta figurines and a terracotta bust of a female figure placed within a small shrine; h. some natural astragali; h. grooming implements (two iron choppers, a bronze mirror, lead and iron grooming tools, a shell palette(?) and a scallop shell (*pecten jacobaeus*) accompanied by a bronze spatula, some of which appear to have been placed inside a small wooden box. Finally, also found was a small black-glazed skyphos (Py 7304) containing a white pigment (psimythion) on its interior, and near it, another shell palette(?). The two shell palettes(?) seem to preserve traces of pigments.

Date: late 4th century BC.

Bessios M, Πιερίδων Στέφανος: Πύδνα, Μεθώνη και οι αρχαιότητες της βόρειας Πιερίας (επιμ. Γ. Τζιφόπουλος), Κατερίνη 2010, σελ. 242-3. In Greek.

**Sample 9A**

Provenance: Pit Grave 21, Makriyialos, N Cemetery of Pydna, Plot 481. Date of excavation: 2009. Excavators M Bessios and Athena Athanassiadou.

A pit grave oriented along the N-S axis. It was cut into the fill of an earlier grave, so its boundaries could not be precisely determined, and part of the skeleton was disturbed during excavation. The deceased was a young individual, positioned with the skull towards the north. A silver coin, found in a poor state of preservation, had been placed in the mouth. Grave goods were concentrated near the lower limbs; these consisted primarily of pottery vessels, but also included a lead vessel found in fragments, containing traces of white lead pigment.

The vessels consisted primary of ceramic vessels but also a lead vessel in a fragmentary state of preservation containing psimythion. The ceramic vessels were classified as follows: a. two unpainted skyphidia (placed one inside the other); b. an aryballoid lekythion featuring a red-figure palmette; c. a black-glazed bolsal-type skyphos d. a poorly fired vessel found in fragments. Further, a black-glazed skyphidion was recovered during the excavation of the tomb, though its original position could not be determined. Finally, two bronze shoe rings were found next to the right foot, while a lead sling bullet bearing an inscription (MEP NA) was discovered inside one of the unpainted skyphidia.

Date: second half of the 4th century BC

**Sample 13A**

Provenance: Pit Grave 751, Makriyialos, N Cemetery of Pydna, Plot 951. Date of excavation: 1998. Excavators M Bessios and Athena Athanassiadou.

This is a pit grave oriented along an east-west axis. It contained the burial of a woman; her skull laid at the eastern end. The grave had been looted. The tomb robbers had disturbed only the area around the skull, with the latter having been pushed next to the left arm. The lower jaw had rolled down toward the chest; a bronze coin minted under Alexander III, originally placed in the deceased's mouth, was found at that same spot.

On the left arm there was placed an iron fibula of the Illyrian type, while two bronze shoe rings were discovered at each foot. A small black-glazed skyphos had been placed beside the deceased, and a white pellet-shaped psimythion was found nearby.

Date: second half of the 4th century BC.

**Note 3:**

Table S 1: Sample 9A. Regression coefficient for multiple regression of the elements with the phases. R2 column reports the highest regression coefficient for the various linear combinations of the phases listed in the column 'Equations'.

| Sample 9A | | |
|---|---|---|
| **Element** | **Equations** | **R2** |
| Pb | Pb = 1.0875*Cerussite + 104.1895 ($R^2$ = 0.6393)<br>Pb = 11.9755*Calcite + 177.1719 ($R^2$ = 0.4538)<br>Pb = -0.0050*Quartz + 586.2750 ($R^2$ = 0.0004)<br>Pb = 1.0352*Cerussite + 0.8283*Calcite + 99.1307 ($R^2$ = 0.6398)<br>Pb = 1.0874*Cerussite + -0.0009*Quartz + 104.3298 ($R^2$ = 0.6393) | 0.6398 |
| | Pb = 11.9735*Calcite + -0.0012*Quartz + 177.3818 ($R^2$ = 0.4538)<br>**Pb = 1.0351*Cerussite + 0.8273*Calcite + -0.0008*Quartz + 99.2670 ($R^2$ = 0.6398)** | |
| Cu | Cu = 0.0030*Cerussite + 1.7443 ($R^2$ = 0.3760)<br>Cu = 0.0325*Calcite + 2.0320 ($R^2$ = 0.2613)<br>Cu = -0.0000*Quartz + 3.3317 ($R^2$ = 0.0005)<br>Cu = 0.0028*Cerussite + 0.0030*Calcite + 1.7236 ($R^2$ = 0.3765)<br>Cu = 0.0030*Cerussite + -0.0000*Quartz + 1.7454 ($R^2$ = 0.3760)<br>Cu = 0.0325*Calcite + -0.0000*Quartz + 2.0335 ($R^2$ = 0.2613)<br>**Cu = 0.0028*Cerussite + 0.0030*Calcite + -0.0000*Quartz + 1.7247 ($R^2$ = 0.3765)** | 0.3765 |
| Fe | Fe = -0.0075*Cerussite + 23.7092 ($R^2$ = 0.0205)<br>Fe = -0.0892*Calcite + 23.4108 ($R^2$ = 0.0168)<br>Fe = 0.0002*Quartz + 20.2835 ($R^2$ = 0.0005)<br>Fe = -0.0060*Cerussite + -0.0247*Calcite + 23.8807 ($R^2$ = 0.0210)<br>Fe = -0.0075*Cerussite + 0.0002*Quartz + 23.6765 ($R^2$ = 0.0209)<br>Fe = -0.0889*Calcite + 0.0002*Quartz + 23.3771 ($R^2$ = 0.0172)<br>Fe = -0.0060*Cerussite + -0.0245*Calcite + 0.0002*Quartz + 23.8468 ($R^2$ = 0.0214) | 0.0005<br>No correlation |
| Ti | Ti = 0.0000*Cerussite + 3.0350 ($R^2$ = 0.0000)<br>Ti = 0.0046*Calcite + 2.8451 ($R^2$ = 0.0003)<br>Ti = -0.0000*Quartz + 3.0558 ($R^2$ = 0.0000)<br>Ti = -0.0006*Cerussite + 0.0112*Calcite + 2.9395 ($R^2$ = 0.0007)<br>Ti = 0.0000*Cerussite + -0.0000*Quartz + 3.0403 ($R^2$ = 0.0000)<br>Ti = 0.0045*Calcite + -0.0000*Quartz + 2.8499 ($R^2$ = 0.0003)<br>Ti = -0.0006*Cerussite + 0.0111*Calcite + -0.0000*Quartz + 2.9448 ($R^2$ = 0.0008) | 0.0008<br>No correlation |
| Cr | Cr = 0.0015*Cerussite + 1.9113 ($R^2$ = 0.1924)<br>Cr = 0.0167*Calcite + 2.0188 ($R^2$ = 0.1400)<br>Cr = -0.0000*Quartz + 2.6333 ($R^2$ = 0.0005)<br>Cr = 0.0014*Cerussite + 0.0019*Calcite + 1.8985 ($R^2$ = 0.1932)<br>Cr = 0.0015*Cerussite + -0.0000*Quartz + 1.9128 ($R^2$ = 0.1926)<br>Cr = 0.0167*Calcite + -0.0000*Quartz + 2.0204 ($R^2$ = 0.1402)<br>**Cr = 0.0014*Cerussite + 0.0019*Calcite + -0.0000*Quartz + 1.9000 ($R^2$ = 0.1934)** | 0.1932 |
| Ca | Ca = -0.0011*Cerussite + 5.1330 ($R^2$ = 0.0113)<br>Ca = -0.0066*Calcite + 4.8541 ($R^2$ = 0.0026)<br>Ca = -0.0000*Quartz + 4.6047 ($R^2$ = 0.0003) | No correlation found |
| | Ca = -0.0019*Cerussite + 0.0130*Calcite + 5.0409 ($R^2$ = 0.0150)<br>Ca = -0.0011*Cerussite + -0.0000*Quartz + 5.1404 ($R^2$ = 0.0118)<br>Ca = -0.0066*Calcite + -0.0000*Quartz + 4.8612 ($R^2$ = 0.0029)<br>Ca = -0.0019*Cerussite + 0.0129*Calcite + -0.0000*Quartz + 5.0484 ($R^2$ = 0.0153) | |
| Ag | Ag = 0.0012*Cerussite + 1.5560 ($R^2$ = 0.1896)<br>Ag = 0.0126*Calcite + 1.6660 ($R^2$ = 0.1319)<br>Ag = 0.0001*Quartz + 2.1552 ($R^2$ = 0.0003)<br>Ag = 0.0011*Cerussite + 0.0005*Calcite + 1.5513 ($R^2$ = 0.1900)<br>Ag = 0.0012*Cerussite + 0.0001*Quartz + 1.5508 ($R^2$ = 0.1901)<br>Ag = 0.0126*Calcite + 0.0001*Quartz + 1.6607 ($R^2$ = 0.1325)<br>**Ag = 0.0011*Cerussite + 0.0005*Calcite + 0.0001*Quartz + 1.5460 ($R^2$ = 0.1906)** | 0.1906 |